\documentclass[11pt,twoside]{article}

\usepackage{asp2006arxiv}
\usepackage{epsf}

\begin{document}
\title{Dark matter millilensing and VSOP-2}   
\author{Kaj Wiik,\altaffilmark{1} Erik Zackrisson,\altaffilmark{1,}\altaffilmark{2,}\altaffilmark{3}
  and Teresa Riehm\altaffilmark{2}} 

\altaffiltext{1}{Tuorla Observatory, University of Turku,%
  V\"ais\"al\"antie 20, FI-21500 Piikki\"o, Finland}%
\altaffiltext{2}{Stockholm Observatory, AlbaNova University Center,%
  106 91 Stockholm, Sweden}%
\altaffiltext{3}{Department of Astronomy and Space Physics, Box 515,%
  751 20 Uppsala, Sweden} 

\begin{abstract}
  According to the cold dark matter scenario, a large number of dark
  subhalos should be located within the halo of each Milky-way sized
  galaxy. One promising possibility for detecting such subhalos is to
  try to observe their gravitational lensing effects on background
  sources. Dark matter subhalos in the $10^6$ -- $10^{10}$ M$_{\odot}$
  mass range should cause strong gravitational lensing on the
  (sub)milliarcsecond scales, which can be observed only using
  space VLBI. We study the feasibility of a
  strong-lensing detection of dark subhalos by deriving the image
  separations expected for density profiles favoured by current
  simulations and comparing it to the angular resolution of both
  existing and upcoming observational facilities. We show that the
  detection of subhalos is likely much more difficult than suggested
  in previous studies, due to the smaller image separations predicted
  for subhalo density profiles more realistic than the singular
  isothermal sphere models often adopted.
\end{abstract}

\section{Introduction}

The quest to unravel the nature of the dark matter, estimated to
contribute around 23\% to the energy density of the Universe (e.g.
\citealt{2006astro.ph..3449S}), remains one of the most important
tasks of modern cosmology. While the cold dark matter (CDM) scenario
-- in which the dark matter particles are assumed to be
non-relativistic at the epoch of decoupling and to interact
predominantly through gravity -- has been very successful in
explaining the formation of large-scale structures in the Universe
(see e.g. \citealt{2003NuPhS.124....3P} for a review), its predictions
on scales of individual galaxies have not yet been confirmed in any
convincing way.

In particular, CDM predicts a large number of subhalos, typically
accounting for $\simeq$ 5-10\% of the total mass of a galaxy-sized CDM
halo. However, these subhalos do not appear to correspond to luminous
structures, as the Milky Way would then be surrounded by a factor of
10--100 more satellite galaxies than observed, provided that each
subhalo hosts a luminous dwarf galaxy (\citealt{1999ApJ...524L..19M}). One
way out of this dilemma is to assume that most of these subhalos
correspond to so-called dark galaxies (\citealt{2002MNRAS.336..541V}),
i.e. objects which either do not contain baryons or in which the
baryons formed very few stars. While a number of very faint
missing-satellite candidates have recently been detected
(\citealt{2006ApJ...640..270S,2006ApJ...650L..41Z}), it is still far
from clear that these exist in sufficient numbers to account for the
subhalos predicted by CDM (e.g. \citealt{2007ApJ...670..313S}).

Since several other CDM predictions of
dark halo properties have recently been called into question (see
e.g. \citealt{2006A&A...452..857Z} and references therein), it is of
paramount importance to investigate whether these CDM subhalo
predictions really hold true.

While dark matter cannot be seen directly in current telescopes, its
presence can be inferred from the gravitational lensing effects that
spatially clustered dark matter will have on background light
sources. If a sufficiently dense and massive dark object happens to be
located along the line of sight to some distant astronomical object --
in most cases a quasar -- the light may reach the observer along
several different paths, thereby producing multiple images in the sky
of a single light source. One tell-tale signature of dark matter
subhalos in the $10^{6}$--$10^{10}\ M_\odot$ mass range could be
gravitational millilensing (sometimes also referred to as
mesolensing), i.e. image splitting at a characteristic separation of
milliarcseconds (e.g. \citealt{1992ApJ...397L...1W,2003PASJ...55.1059Y}). 

Based on a null detection of millilensing in a sample of 300 quasars
observed with the VLBI, Wilkinson et
al. (\citeyear{2001PhRvL..86..584W}) demonstrated that the vast
majority of quasars do {\em not} show any signs of millilensing in the
angular range 1.5 -- 50 milliarcseconds, and were able to impose upper
limits of $\Omega<0.01$ on the cosmological density of dark point-mass
objects in the $10^{6}$--$10^{8}\ M_\odot$ mass range. Unfortunately,
this constraint is still insufficient to rule out the subhalos
predicted by CDM, since their lensing optical depth is expected to be
at least one order of magnitude lower.

\section{Subhalo millilensing}
To put the CDM subhalo predictions to the test, Yonehara et
al. (\citeyear{2003PASJ...55.1059Y}) suggested that one should target
quasars which are already known to be gravitationally lensed by
galaxies on arcsecond scales, as one can then be sure that there is a
massive halo well-aligned with the line of sight. In this case there
should be a significant probability (optical depth $\tau \simeq$ 0.01
-- 0.1) of detecting image splitting by subhalos at scales of
milliarcseconds (millilensing). Indeed, millilensing
has long been suspected to be the cause of the flux ratio anomalies
seen in such systems
(e.g. \citealt*{1998MNRAS.295..587M,2004ApJ...610...69K}).  Subhalo
millilensing has also been advocated as an explanation for strange
bending angles of radio jets in these multiply-imaged quasars
(\citealt{2002ApJ...580..696M}) and for image positions which smooth
halo models seem unable to account for
(\citealt{2004MNRAS.350..949B}). In cases where the quasar images are
sufficiently resolved, the technique proposed by Yonehara et al. would
not only allow the {\em detection} of subhalos, but will also impose
interesting constraints on their internal density profiles
(\citealt{2005ApJ...634...77I}).

\section{Subhalo density profiles and image separation}

The proposed observations to detect image splitting
(e.g. \citealt{2003PASJ...55.1059Y}) assume that the image
separations of subhalo lenses are similar to those produced by a
singular isothermal sphere (SIS). Unfortunately this assumption is
difficult to justify because theoretical arguments, simulations, and
observations do not favour this form of density profile for dark
matter halos in the relevant mass range. We have studied the
feasibility of strong-lensing detection of dark subhalos by deriving
the image separations expected for (more realistic) density profiles
favoured by current simulations (\citealt{erik_submitted}).

\begin{figure}
\plottwo{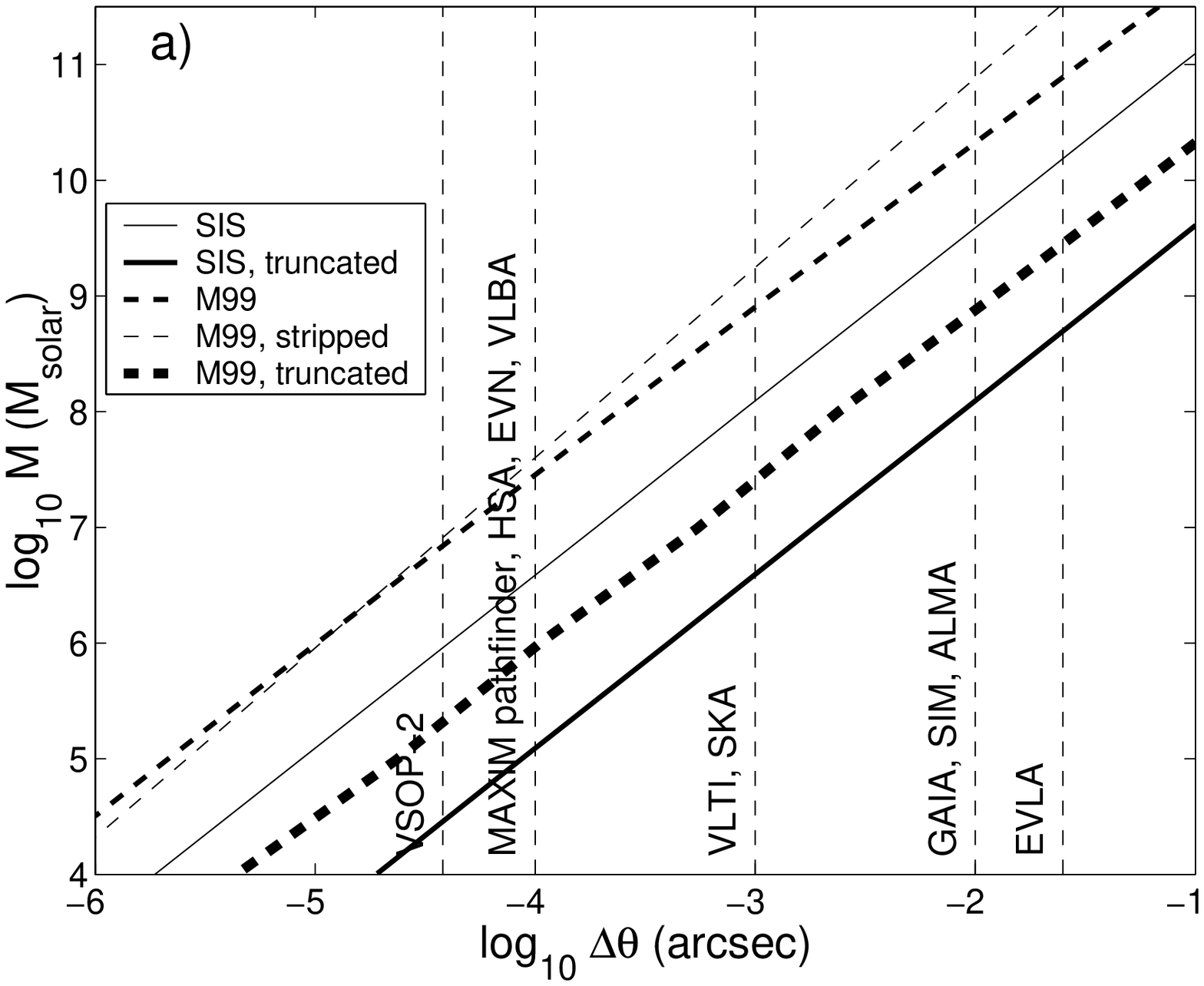}{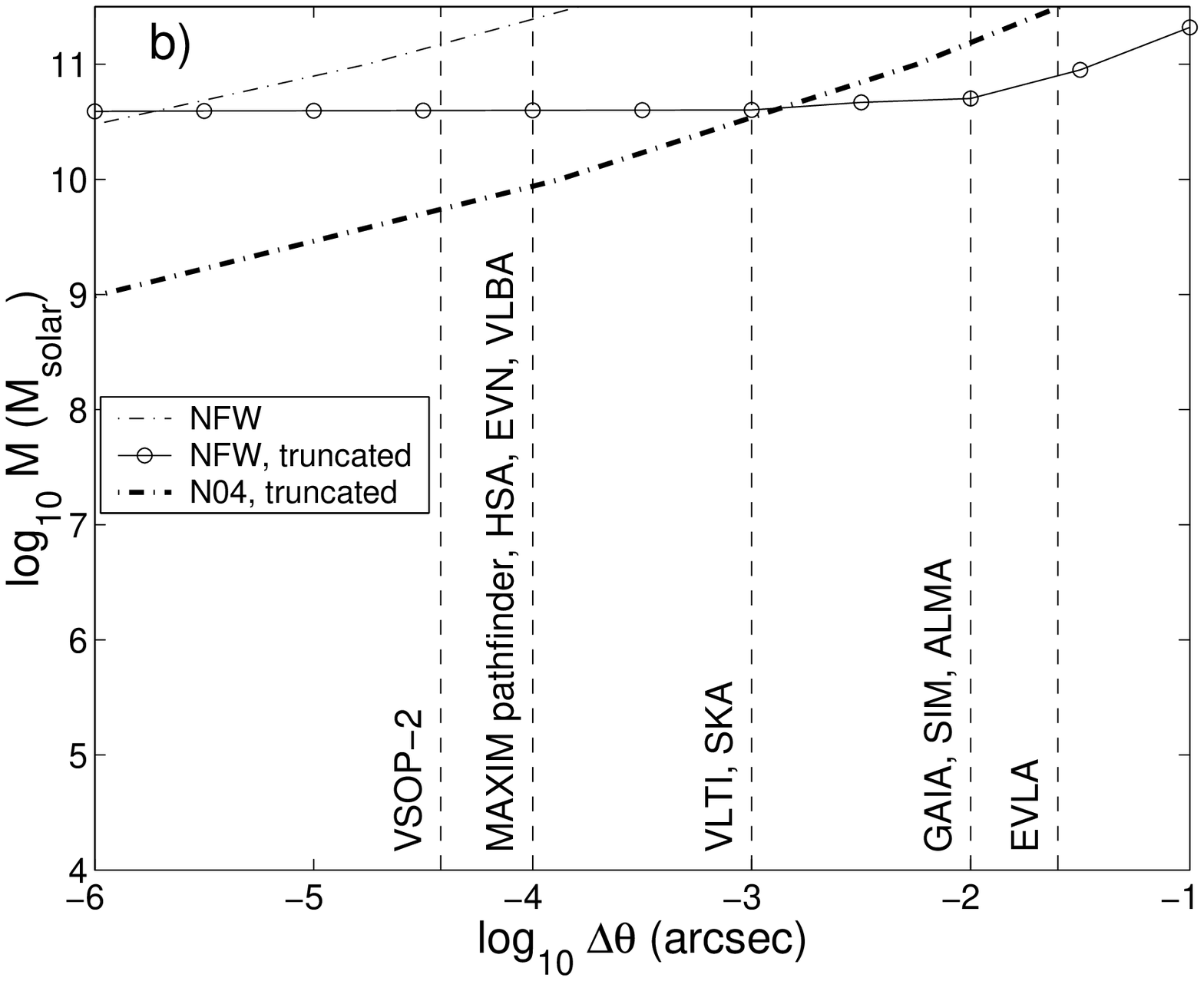}
\caption{Image separation $\Delta\theta$ versus subhalo mass for those
  density profiles that give rise to image separations on scales
  larger than microarcseconds. SIS (thin solid), truncated SIS (thick
  solid), NFW (thin dash-dotted), truncated NFW (solid with circles),
  truncated N04 (thick dash-dotted), M99
  (medium dashed), stripped M99 (thin dashed) and truncated M99 (thick
  dashed). The angular resolution of a number of existing and planned
  observational facilities has been indicated by vertical dashed
  lines.\label{sepfig}}
\end{figure}

In Fig.~\ref{sepfig} we have plotted the calculated image separation
for four density profiles that exceeded one microarcsecond. These are
SIS, NFW (\citealt{1996ApJ...462..563N}), M99
(\citealt{1999ApJ...524L..19M}), and N04
(\citealt{2004MNRAS.349.1039N}). These profiles are all based on
simulations for relatively isolated halos. When such objects are
accreted by more massive halos and become subhalos, substantial mass
loss occurs, preferentially from the outer regions of the
satellites. To account for this, the models are truncated or gradually
stripped. For details, see \cite{erik_submitted}.

\section{Discussion}

It is clear from Fig.~\ref{sepfig} that detection of subhalos is
likely considerably more difficult than suggested by previous
studies. We stress however that these simulations do not necessarily
represent the final word on this issue. If the subhalos have central
density slopes steeper than $\rho\propto r^{-1}$ (e.g. of M99 type;
with $\rho\propto r^{-1.5}$) , these would give rise to image
separations that could be resolved even with current telescopes.

\cite{2005ApJ...634...77I} proposed to search already known
multiply-imaged quasars (macrolensed objects) for substructures. If
the true density profile of subhalos is shallow and leads to small
image separations, and hence small optical depths, perhaps the number
of the known suitable lensed AGN is insufficient even for a single
detection. We are going to investigate if the detection probability
would increase by targeting AGN with larger impact parameters, by
trying to estimate the average detectable area of AGN jets using data
from the published surveys and compare that with the predicted optical
depth of the subhalo objects. At lower frequencies the sources are
larger in area but the resolution is worse. Another goal is to find
the optimum frequency for millilensing search with the VSOP-2.

To increase the detection probability, {\em all images} that are
produced by VSOP-2 should be looked at with this effect in mind. This
could be done either after the data has become public or in a key
science program that would produce 'quick look' images using a pipeline
immediately after the data has been correlated.

\acknowledgements EZ acknowledges research grants from the Swedish
Research Council, the Royal Swedish Academy of Sciences and the
Academy of Finland. TR acknowledges support from the HEAC Centre
funded by the Swedish Research Council. KW acknowledges support from
the Jenny and Antti Wihuri foundation.

\end{document}